\documentclass[smallextended]{svjour3}       

\smartqed

\usepackage[centertags]{amsmath}
\usepackage{amsfonts}
\usepackage{amssymb}

\usepackage{color}

\def\Lie{\pounds}
\def\dual{{}^\star\!}
\def\N{\textbf{\scriptsize N}}

\begin{document}

\title{Quasi-local Energy for Spherically Symmetric Spacetimes}



\author{Ming-Fan Wu \and Chiang-Mei Chen \and Jian-Liang Liu \and James M. Nester}


\institute{M.-F. Wu \and C.-M. Chen \and J.-L. Liu  \at
              Department of Physics \& Center for Mathematics and Theoretical Physics, National Central University, Chungli 320, Taiwan \\
              \email{93222036@cc.ncu.edu.tw}
           \and
           C.-M. Chen \at
              \email{cmchen@phy.ncu.edu.tw}
           \and
           J.-L. Liu \at
              \email{liujl@phy.ncu.edu.tw}
           \and
           J. M. Nester \at
              Department of Physics \& Center for Mathematics and Theoretical Physics, National Central University, Chungli 320, Taiwan \\
              Graduate Institute of Astronomy, National Central University, Chungli 320, Taiwan \\
              \email{nester@phy.ncu.edu.tw}
}

\date{Received: date / Accepted: date}

\maketitle

\begin{abstract}
We present two complementary approaches for determining the reference for the covariant Hamiltonian boundary term quasi-local energy and test them on spherically symmetric spacetimes.
On the one hand, we isometrically match the 2-surface and extremize the energy. This can be done in two ways, which we call programs I (without constraint) and II (with additional constraints).
On the other hand, we match the orthonormal 4-frames of the dynamic and the reference spacetimes. Then, if we further specify the observer by requiring the reference displacement to be the timelike Killing vector of the reference, the result is the same as program I, and the energy can be positive, zero, or even negative. If, instead, we require that the Lie derivatives of the two-area along the displacement vector in both the dynamic and reference spacetimes to be the same, the result is the same as program II, and it satisfies the usual criteria: the energies are non-negative and vanish only for Minkowski (or anti-de Sitter) spacetime.


\end{abstract}

\maketitle

\section{Introduction}
General relativity is no doubt one of the most important and beautiful theories of the 20th century. In this theory, Einstein's happiest thought, the equivalence principle, plays a most essential role. But somewhat ironically it also leads us into difficulties in having a proper definition for the energy density of gravitating systems, therefore we still do not have a satisfactory description for gravitational energy. The modern concept is that gravitational energy should be non-local, more precisely {\it quasi-local}, i.e., it should be associated with a closed two-surface (for a comprehensive review see~\cite{Szabados:2004vb}). Here we consider one proposal based on the covariant Hamiltonian formalism~\cite{Nester91} wherein the quasi-local energy is determined by the Hamiltonian boundary term. For a specific spacetime displacement vector field on the boundary of a region (which can be associated with the observer), the quasi-local energy---defined as the value of the Hamiltonian boundary term---depends not only on the dynamical values of the fields on the boundary but also on the choice of reference values for these fields. Thus a principal issue in this formalism is the proper choice of reference spacetime for a given observer, which is equivalent to finding a suitable embedding of a neighborhood of the boundary 2-surface into a suitable reference space. 
Of particular interest here is the use of the quasi-local energy value to find the embedding variables.

It is generally accepted that the total energy for an asymptotically flat gravitating system should be non-negative and should vanish only for Minkowski space (this is required for stability, see, e.g.,~\cite{Brill:1968ca}; for proofs of this property for GR see~\cite{Schoen:1979zz,Witten:1981mf}). In view of this these properties have also been regarded by many as being desirable for a good quasi-local energy~\cite{Szabados:2004vb,Liu:2003bx,Liu:2004dc,Wang:2008jy,WaYa2008}.

Following up on one approach used in~\cite{Liuthes,Liu:2011jha}, recently we proposed an energy-extremization program~\cite{Chen:2009zd,Wu:2011wk} to determine the reference, and found that it works well in spherically symmetric spacetimes. We have successfully obtained the quasi-local energy for any given displacement vector. However, since the program was designed to produce the minimum energy---the program chooses one specific embedding as the ground state such that the energy produced is the minimum---there is no guarantee of getting non-negative energies.  We find the stability arguments for positive energy compelling for asymptotically flat regions approaching equilibrium.  As discussed in some detail in~\cite{Liu:2011jha,Chen:2009zd,Wu:2011wk,Nester:2008xd}, we do not see why a good measure of quasi-local energy for regions that are highly dynamic and/or not asymptotically flat must necessarily have positive energy.

Although for our agenda, that is, to find the ``best'' choice of reference for our Hamiltonian boundary term, negative results can be accepted, we still wonder under what conditions one can get the energies that satisfy the usual criteria, i.e., it should be non-negative and vanishes only for Minkowski (or (anti)-de Sitter) spacetime.  Our idea is that for any observer in the dynamic spacetime one should pick a comparable reference observer as the ground state. Therefore we should minimize the energy under some suitable constraints.

There are some common themes and aspects of the work here that are similar to those in the recent quasi-local work of Wang and Yau~\cite{Wang:2008jy,WaYa2008}.  Of course their embedding construction is much more general and intricate than our spherically symmetric case. In~\cite{WaYa2008} they have used an optimal embedding, a strategy that we will use here too.  There are also significant differences.  First and foremost they have a quite general positive energy proof for their energy.  In the present paper, our spacetime vector is generally allowed to be picked by hand, consequently the embedding variables found here usually depend on this vector. We have considered here embeddings into a background anti-de Sitter spacetime not just Minkowski.  Our embeddings are not just 2-surface isometric embeddings into Minkowski space, they match on the boundary the 4D dynamic and reference metrics.
Also it should be noted that, although there are certain similarities, there are also important differences between the quasi-local expressions we have developed and those based on the Brown and York~\cite{Brown:1992br} Hamilton-Jacobi approach, including in particular the expressions used by Wang and Yau.




In this paper we propose a modified energy-extremization program, in which we minimize the energy under both the following constraints:

$\textbf{C1}. \quad \Lie_{\N} (\vartheta^2 \wedge \vartheta^3) = \Lie_{\N} (\bar{\vartheta}^2 \wedge \bar{\vartheta}^3)$,

$\textbf{C2}. \quad \vartheta^0 = \bar{\vartheta}^0, \quad \textbf{N} = l e_0 = l \bar{e}_0, \quad l^2 = - g(\textbf{N}, \textbf{N})$,

\noindent
where $\Lie_{\N}$ means the Lie derivative along the displacement vector $\textbf{N}$, and $\vartheta^{\mu}(\bar{\vartheta}^{\mu})$ means the orthonormal coframe of the dynamic(reference) spacetime. The geometric meaning of these constraints is clear. The constraint \textbf{C1} requires the time evolution of the two-area for both the dynamic and reference observers to be the same on the boundary. It should be kept in mind that throughout this paper we always impose the two-surface isometric embedding condition, so the two-areas at some constant time are equal too, i.e., $\vartheta^2 \wedge \vartheta^3 = r^2 \sin\theta d\theta \wedge d\phi = \bar{\vartheta}^2 \wedge \bar{\vartheta}^3$. Since the quasi-local energy is defined on the two-boundary, it is natural to pick a reference observer in this way so we will not compare the energy for a expanding two-area with a shrinking one. The constraint \textbf{C2} requires the displacement vector $\textbf{N}$ to be the normal of the spacelike hypersurfaces $\vartheta^0 = 0$ and $\bar{\vartheta}^0 = 0$, and it has the same magnitude for both dynamic and reference spacetimes. In other words, we synchronize the two clocks and foliate both spacetimes in the same manner such that the flow of time is orthogonal to the hypersurface at the boundary 2-surface. These constraints determine what we mean by comparable reference observers both geometrically and physically. Applied to the spherically symmetric spacetimes, the results satisfy the usual criteria. The only exception is that for a small region inside the inner horizon of the Reissner-Nordstr\"{o}m spacetime the energy is negative. This negative result is not so strange as this is a rather odd region where the gravitational force becomes repulsive.

In addition to the produced quasi-local energies, it is also very interesting and important to investigate the references determined by our energy extremization programs. It turns out they are the embeddings that exactly match the whole orthonormal frames right on the two-sphere boundary. This result motivates us to propose another complementary approach to the whole problem, that is, we propose that the optimal choice of the reference is the one---at least for spherically symmetric cases---which matches the 4D metric on the boundary. Applying this idea, we find the resultant energy still cannot be uniquely determined. This can be understood: in order to get a meaningful quasi-local energy we also need to specify a reference observer to compare with, so we should further impose some physical conditions. There are two obviously physically meaningful choices we can think of. First, it is reasonable to require the reference displacement vector to be (proportional to) the timelike Killing vector, i.e. $\textbf{N} = N^T \partial_T$. Second, instead of always picking the reference timelike Killing vector as our displacement vector, we can also pick a comparable reference observer as the ground state. By that we mean requiring $\Lie_{\N} (\vartheta^2 \wedge \vartheta^3) = \Lie_{\N} (\bar{\vartheta}^2 \wedge \bar{\vartheta}^3)$ on the boundary. As we are going to show, these two choices are equivalent to energy extremization programs I and II respectively. Therefore the pictures of the quasi-local energy for spherically symmetric spacetimes are quite complete and satisfactory. On the one hand, we know the geometric meaning of the energy extremization programs. On the other hand, the most optimized references for spherically symmetric spacetimes, i.e. those that match the geometry on the boundary, produce the extremum quasi-local energies.

In the following section we briefly review the quasi-local expression. Then apply both energy-extremization programs to spherically symmetric spacetimes in section 3. In section 4 we turn to the isometric matching approach and show that the results are compatible with the energy-extremization programs. A discussion section concludes this paper.

\section{Quasi-local Expression}
In this section we first briefly sketch the preferred quasi-local energy-momentum Hamiltonian boundary term expression for Einstein's gravity theory which was derived using the covariant Hamiltonian formalism. The details of the derivation and related discussions can be found in~\cite{Chen:1994qg,Chen:1998aw,Chang:1998wj,Chen:2000xw,Chen:2005hwa,Nester:2008} (for some additional developments along similar lines see~\cite{Anco:2001gk,Anco:2001gm}).

For a theory which is invariant under diffeomorphisms, in particular infinitesimal displacements along some vector field $\textbf{N}$, the Hamiltonian which generates the dynamical evolution of a spatial region along such a vector field is given by the integral over the region of a suitable Hamiltonian 3-form which takes the form
\begin{equation} \label{H+dB}
{\cal H}(\textbf{N}) =: N^\mu {\cal H}_\mu + d {\cal B}(\textbf{N}).
\end{equation}
As a consequence of {\em local} diffeomorphism invariance (i.e., a symmetry for non-constant $N^\mu$), the density part ${\cal H}_\mu$ must be proportional to certain 4-covariant field equations, hence it vanishes ``on shell''. Consequently the value of the conserved quantity associated with a local displacement $\textbf{N}$ and a spatial region $\Sigma$ is determined entirely by a 2-surface integral over the boundary of the region:
\begin{equation} \label{EN}
E(\textbf{N}, \Sigma) := \int_\Sigma {\cal H}(\textbf{N}) = \oint_{\partial\Sigma} {\cal B}(\textbf{N}).
\end{equation}
The value is thus {\em quasi-local} (depending only on the values of the fields on the boundary).  For {\em any} choice of $\textbf{N}$ this expression defines a conserved quasi-local quantity. Several particular boundary terms were identified, each is associated with a distinct type of boundary condition.  In~\cite{Chen:2005hwa} a ``preferred boundary term'' (it has a certain covariant property, directly gives the Bondi energy flux, and has a positive total energy proof) for GR was identified:
\begin{equation} \label{expB}
{\cal B}(\textbf{N}) = \frac{1}{16\pi} (\Delta\omega^{\alpha}{}_{\beta} \wedge i_{\N} \eta_{\alpha}{}^{\beta} + \bar D_{\beta} N^\alpha \Delta \eta_{\alpha}{}^\beta),
\end{equation}
where $\omega^{\alpha\beta} = \omega^{[\alpha\beta]}$ is the {\em connection one-form}, $\eta^{\alpha\beta} := \dual (\vartheta^\alpha \wedge \vartheta^\beta)$ is a 2-form depending on the {\em coframe} $\vartheta^\alpha$, $\Delta$ indicates the difference between the dynamic and reference values, and $\bar D_{\beta}$ is the covariant derivative using the reference connection. The reference values can be determined by pullback from an embedding of a neighborhood of the boundary into a suitable reference space. Now we can use this expression to calculate the gravitational energy in general relativity.

\section{Energy-Extremization Program}
In this section, we first briefly review the results (and extend them using a generalization from the Schwarzschild to the Reissner-Nordstr\"{o}m--(anti)-de Sitter (RN--(A)dS) spacetime) obtained in our earlier papers~\cite{Chen:2009zd,Wu:2011wk}, where we extremized the energy with respect to the embedding variables without any further constraint other than the two-surface isometric embedding condition. Then we present our modified program where we extremize the energy under some additional suitable constraints. For convenience we call them programs I (without constraint) and II (with constraints). Note again that throughout this paper we always impose the two-surface isometric embedding condition.

\bigskip
\noindent {\bf Program I}: \\
Extremize the energy with respect to the embedding variables without any further constraint.

\bigskip
We first consider the static spherically symmetric case, more specifically, we consider the RN--(A)dS spacetime,
\begin{equation}
ds^2 = -A dt^2 + A^{-1} dr^2 + r^2 d\Omega^2, \label{RN-(A)dS}
\end{equation}
where
\begin{equation}
A = 1 - \frac{2m}{r} + \frac{Q^2}{r^2} - \frac{\Lambda}{3}r^2, \qquad d\Omega^2 = d\theta^2 + \sin^2{\theta} d\phi^2.
\end{equation}
Choose the pure (anti)-de Sitter spacetime as the reference,
\begin{equation}
d\bar{s}^2 = -B dT^2 + B^{-1} dR^2 + R^2 d\Omega^2, \label{(A)dSref}
\end{equation}
where $B = 1 - \frac{\Lambda}{3} R^2$. A legitimate approach for the spherically symmetric case is to assume $T = T(t,r), R = R(t,r), \Theta = \theta, \Phi = \phi$ along with $R_0 := R(t_0, r_0) = r_0$; this symmetrically embeds a neighborhood of the two-sphere boundary $S$ at $(t_0, r_0)$ into the (A)dS reference such that the two-sphere boundary is embedded isometrically.

Because the spacetime is spherically symmetric, we expect that the quasi-local energy can be completely determined by the first term of the expression~(\ref{expB}).
In earlier works, via both explicit calculation and theoretical analysis~\cite{Chen:1998aw,Hecht:1993qf,HN96,Nester:2004xm,Nester:2004}, it was found that the second term in our quasi-local expression makes important contributions to the value of the angular momentum and the center-of-mass moment.  This term is not expected to make a contribution to the energy of spherically symmetric systems, and we have not noticed any case where it does.  Here in this section we will assume for simplicity that this holds and will proceed to determine the reference using just the first term in  our expression.  It will turn out that the reference determined by this procedure indeed does guarantee that the second term vanishes, so this procedure is self-consistent.  In the next section we give a further confirmation: we will obtain the same results from a different assumption which definitely kills the second term.

Now, using just the first term, 
from~(\ref{EN}, \ref{expB}) the quasi-local energy of such a spacetime in terms of the displacement vector
\begin{equation}
\textbf{N} = N^t \partial_t + N^r \partial_r = N^T \partial_T + N^R \partial_R,
\end{equation}
and embedding variables $(T_t, T_r, R_t, R_r)$ can be written as
\begin{equation} \label{Energy}
E = \frac{r}{2} \left( B X N^T + A N^t R_r + A^{-1} N^r R_t - 2 A N^t \right),
\end{equation}
where $X^{-1} = T_t R_r - T_r R_t$.
By extremizing the energy~(\ref{Energy}) with respect to the four embedding variables we get
\begin{eqnarray}
\frac{\partial E}{\partial T_t} = 0 &\qquad \Rightarrow \qquad& B X^2 T_r N^R = 0, \label{ETt}
\\
\frac{\partial E}{\partial T_r} = 0 &\Rightarrow& B X^2 T_t N^R = 0, \label{ETr}
\\
\frac{\partial E}{\partial R_t} = 0 &\Rightarrow& B X^2 T_r N^T + A^{-1} N^r = 0, \label{ERt}
\\
\frac{\partial E}{\partial R_r} = 0 &\Rightarrow& B X^2 T_t N^T - A N^t = 0. \label{ERr}
\end{eqnarray}
Note that the first two equations are equivalent (since we do not want both $T_t$ and $T_r$ to vanish), so we only have three independent equations.
Exploiting these equations we can get
\begin{equation} \label{RtRr}
N^R = 0, \quad X N^T = \frac{N^t}{R_r}, \quad R_r = \sqrt{\frac{-B}{g(\textbf{N}, \textbf{N})}} N^t, \quad R_t = -\sqrt{\frac{-B}{g(\textbf{N}, \textbf{N})}} N^r,
\end{equation}
so that the energy obtained is
\begin{equation} \label{energy SA}
E_{\rm I} = r \left( \sqrt{-g(\textbf{N}, \textbf{N}) B} - A N^t \right).
\end{equation}

For a static observer at a distance $r$ (note: such observers exist only for values of $r$ such that $A$ is positive)
\begin{equation}
\textbf{N}_{\rm static} = \frac{1}{\sqrt{A}} \partial_t,
\end{equation}
the energy measured is\footnote{This is meaningful only for ranges of $r$ where $B$ and $A$ are positive.  A similar remark  applies to many of our later expressions. We will often not explicitly mention the appropriate domain of validity; it can easily be determined by a detailed examination of the expressions.}
\begin{equation}
E_{\rm I}(\textbf{N}_{\rm static}) = \frac{2 m - Q^2/r}{\sqrt{B} + \sqrt{A}}.
\end{equation}
For Schwarzschild this reduces to a standard result, $E = r(1 - \sqrt{1 - 2m/r})$, first found by Brown and York~\cite{Brown:1992br}.

For a radial geodesic observer who falls initially with velocity $v_0$ from a constant distance $r = a > 2m$ in Schwarzschild spacetime, i.e., $Q = 0, \Lambda = 0$,
\begin{equation}
N^t = \frac{1}{1 - 2 m/r} \sqrt{\frac{1 - 2m/a}{1 - v_0^2}}, \qquad g(\textbf{N}, \textbf{N}) = -1,
\end{equation}
the energy measured according to our program is
\begin{equation}
E_{\rm I}(\textbf{N}_{\rm geo}) = r \left( 1 - \sqrt{\frac{1 - 2m/a}{1 - v_0^2}} \right);
\end{equation}
(values which were also found by other techniques, see~\cite{Blau:2007wj,Yu:2008ij}). When the initial velocity $v_0$ is less, equal, or greater than the escape velocity, $\sqrt{2m/a}$, the energy is positive, zero, or negative, respectively. One interesting fact is that the sign of the scalar curvature of the hypersurface orthogonal to this displacement vector, $\textbf{N}_{\rm geo}$, is the same as the energy. This feature is very much like the cosmology case which we are going to see later.

It is obvious that the energy-extremization equations~(\ref{ETt}--\ref{ERr}) cannot uniquely determine the reference, since only three of them are independent. However, by further imposing the condition $g(\textbf{N}, \textbf{N}) = \bar{g}(\textbf{N}, \textbf{N})$ on the boundary the reference determined is unique and kills the second term of the expression~(\ref{expB}).

\bigskip
\noindent {\bf Program II}: \\
Extremize the energy with respect to the embedding variables under both the following constraints.

$\textbf{C1}. \quad \Lie_{\N} (\vartheta^2 \wedge \vartheta^3) = \Lie_{\N} (\bar{\vartheta}^2 \wedge \bar{\vartheta}^3)$,

$\textbf{C2}. \quad \vartheta^0 = \bar{\vartheta}^0, \quad \textbf{N} =l e_0 = l \bar{e}_0, \quad l^2 = -g(\textbf{N}, \textbf{N})$,

\noindent where $\Lie_{\N}$ means the Lie derivative along the displacement vector $\textbf{N}$.

\bigskip
From constraint \textbf{C1} we get
\begin{eqnarray}
&& \Lie_{\N} (\vartheta^2 \wedge \vartheta^3) = 2 r \sin\theta N^r d\theta \wedge d\phi,
\nonumber\\
&& \Lie_{\N} (\bar{\vartheta}^2 \wedge \bar{\vartheta}^3) = 2 r \sin\theta N^R d\theta \wedge d\phi,
\nonumber\\
\Rightarrow \quad && \Lie_{\N} (\vartheta^2 \wedge \vartheta^3) = \Lie_{\N} (\bar{\vartheta}^{2} \wedge \bar{\vartheta}^{3}) \qquad \Rightarrow \qquad N^r = N^R, \label{NrequalNR}
\end{eqnarray}
where the useful formula $\Lie_{\N} = i_{\N} d + d i_{\N}$ is used.  Note that these constraints are imposed only on the boundary, where (because of isometric matching) $R = r$. Then from~(\ref{NrequalNR}) we get the relation between $R_t$ and $R_r$,
\begin{equation} \label{Rt}
N^t R_t + N^r R_r = N^R = N^r \qquad \Rightarrow \qquad R_t = \frac{N^r}{N^t}(1 - R_r).
\end{equation}

In order to apply constraint \textbf{C2}, we first write the coframes in the form
\begin{eqnarray}
\vartheta^0 = a_t dt + a_r dr, &\qquad& \vartheta^1 = b_t dt + b_r dr,
\nonumber\\
\bar{\vartheta}^0 = a_T dT + a_R dR, &\qquad& \bar{\vartheta}^1 = b_T dT + b_R dR. \label{theta01}
\end{eqnarray}
This is quite general since we only consider spherically symmetric cases here; what is being allowed are radial boosts  of the obvious coframes associated with~(\ref{RN-(A)dS}) and~(\ref{(A)dSref}). Note that one can also express the displacement vector using the orthonormal frames
\begin{eqnarray}
\textbf{N} &=& N^t \partial_t + N^r \partial_r = N^T \partial_T + N^R \partial_R
\nonumber\\
&=& N^0 e_0 + N^1 e_1 = N^{\bar{0}} \bar{e}_0 + N^{\bar{1}} \bar{e}_1. \label{NproB}
\end{eqnarray}
Using~(\ref{theta01}) and~(\ref{NproB}) we get:
\begin{eqnarray}
N^0 &=& \vartheta^0(\textbf{N}) = a_t N^t + a_r N^r,
\nonumber\\
N^1 &=& \vartheta^1(\textbf{N}) = b_t N^t + b_r N^r,
\nonumber\\
N^{\bar{0}} &=& \bar{\vartheta}^0(\textbf{N}) = a_T N^T + a_R N^R,
\nonumber\\
N^{\bar{1}} &=& \bar{\vartheta}^1(\textbf{N}) = b_T N^T + b_R N^R.
\nonumber
\end{eqnarray}
Together with
\begin{eqnarray}
&& -A dt^2 + A^{-1} dr^2 = - (\vartheta^0)^2 + (\vartheta^1)^2
\nonumber\\
\Rightarrow \quad && - a_t^2 + b_t^2 = -A, \quad - a_r^2 + b_r^2 = A^{-1}, \quad - a_t a_r + b_t b_r = 0, \label{atbt}
\end{eqnarray}
and
\begin{eqnarray}
&& -B dT^2 + B^{-1} dR^2 = -(\bar{\vartheta}^0)^2 + (\bar{\vartheta}^1)^2
\nonumber\\
\Rightarrow \quad && -a_T^2 + b_T^2 = - B, \quad -a_R^2 + b_R^2 = B^{-1}, \quad -a_T a_R + b_T b_R = 0, \label{aTbT}
\end{eqnarray}
one can work out the following relations:
\begin{eqnarray}
a_t = \frac{A N^0 N^t \mp N^1 N^r}{-g(\textbf{N}, \textbf{N})}, &\qquad& a_r = \frac{-A^{-1} N^0 N^r \pm N^1 N^t}{-g(\textbf{N}, \textbf{N})},
\nonumber\\
b_t = \frac{A N^1 N^t \mp N^0 N^r}{-g(\textbf{N}, \textbf{N})}, & & b_r = \frac{-A^{-1} N^1 N^r \pm N^0 N^t}{-g(\textbf{N}, \textbf{N})},
\nonumber\\
a_T = \frac{B N^{\bar{0}} N^T \mp N^{\bar{1}} N^R}{-g(\textbf{N}, \textbf{N})}, & & a_R = \frac{-B^{-1} N^{\bar{0}} N^R \pm N^{\bar{1}} N^T}{-g(\textbf{N}, \textbf{N})},
\nonumber\\
b_T = \frac{B N^{\bar{1}} N^T \mp N^{\bar{0}} N^R}{-g(\textbf{N}, \textbf{N})}, & & b_R = \frac{-B^{-1} N^{\bar{1}} N^R \pm N^{\bar{0}} N^T}{-g(\textbf{N}, \textbf{N})}. \label{abtr}
\end{eqnarray}
In order to preserve the orientation, one should choose the upper sign. Now by applying constraint \textbf{C2} we get
\begin{eqnarray}
&& \vartheta^0 = a_t dt + a_r dr = a_T dT + a_R dR = \bar{\vartheta}^0
\nonumber\\
\Rightarrow \quad && a_t = a_T T_t + a_R R_t, \quad a_r = a_T T_r + a_R R_r, \label{theta00bar}
\end{eqnarray}
and
\begin{equation}
\textbf{N} = l e_0 = l \bar{e}_0 \quad \Rightarrow \quad N^1 = N^{\bar{1}} = 0. \label{N1N1bar}
\end{equation}
By applying~(\ref{abtr}--\ref{N1N1bar}) we get the following relations
\begin{eqnarray}
&& T_t = \frac{1}{B N^T} (A N^t + B^{-1} N^R R_t),
\nonumber\\
&& T_r = \frac{1}{B N^T} (-A^{-1} N^r + B^{-1} N^R R_r).  \label{TtTr}
\end{eqnarray}
Together with~(\ref{Rt}) then there is now only one independent variable, which we choose to be $R_r$. Extremizing the energy with respect to $R_r$ gives
\begin{equation}
R_r = \frac{-1}{g(\textbf{N}, \textbf{N})} \left( B N^T N^t - A^{-1} (N^r)^2 \right).
\end{equation}
Note that by straightforward calculation one can see that the determined reference kills the second term of~(\ref{expB}). The energy obtained is
\begin{eqnarray}
E_{\rm II} &=& r \left( \sqrt{-g(\textbf{N}, \textbf{N}) B + (N^r)^2} - A N^t \right)
\nonumber\\
&=& \frac{-g(\textbf{N}, \textbf{N}) (2m - Q^2/r)}{\sqrt{-g(\textbf{N}, \textbf{N}) B + (N^r)^2} + A N^t}. \label{energy SB}
\end{eqnarray}
One can see that for any timelike displacement vector this energy is always non-negative except for a small region inside the inner horizon of the RN black hole, $r < Q^2/2m$. The transition is exactly at the turning point where the gravitational force becomes repulsive. For the case $m = Q = 0$ the energy vanishes for all displacement vectors, just as it should, whereas in the case $m = Q = 0$,~(\ref{energy SA}) vanishes only when $N^r = 0$.

For a static observer, the result is the same as program I,
\begin{equation}
E_{\rm II}(\textbf{N}_{\rm static}) = E_{\rm I}(\textbf{N}_{\rm static}).
\end{equation}
For the radial geodesic observer in Schwarzschild spacetime the energy measured is
\begin{equation}
E_{\rm II}(\textbf{N}_{\rm geo}) = \frac{2m}{\sqrt{1 + (N^r)^2} + \sqrt{\frac{1 - 2m/a}{1 - v_0^2}}},
\end{equation}
which is positive for any initial velocity $v_0$.

For the dynamic case, these programs work well too for the Friedman-Lema\^{\i}tre-Robertson-Walker (FLRW) metric. The FLRW metric in the standard spherical coordinate is
\begin{equation}
ds^2 = -dt^2 + \tilde A^2 dr^2 + a^2(t) d\Omega^2, \qquad \tilde A = \frac{a(t)}{\sqrt{1 - k r^2}}.
\end{equation}
Apply program I, the resultant energy obtained from expression~(\ref{expB}) is
\begin{equation}
E_{\rm I} = a r \left( \sqrt{-g(\textbf{N}, \textbf{N})} - \tilde A^{-1} a N^t - \tilde A \dot{a} r N^r \right), \label{energy DA}
\end{equation}
where $\dot{a} = da/dt$.

For a comoving observer this gives a value that was found earlier in~\cite{Chen:2007gx},
\begin{equation}
E_{\rm I}(\partial_t) = \frac{k a r^3}{1 + \sqrt{1 - k r^2}},
\end{equation}
which is positive, zero, and negative for $k = +1, 0, -1$ respectively. An interesting geometrically preferred observer is the one preserving the area. The associated evolution vector is the dual mean curvature vector, known as the Kodama vector.  In this case we have
\begin{equation} \label{Kvec}
\textbf{N}_{\rm Kodama} = \frac{\sqrt{1 - k r^2}}{\sqrt{1 - k r^2 - \dot{a}^2 r^2}} \partial_t - \frac{\dot{a} r}{a} \frac{\sqrt{1 - kr^2}}{\sqrt{1 - k r^2 - \dot{a}^2 r^2}} \partial_r.
\end{equation}
The corresponding energy is
\begin{equation}
E_{\rm I}(\textbf{N}_{\rm Kodama}) = \frac{a r^3 (k + \dot{a}^2)}{1 + \sqrt{1 - k r^2 - \dot{a}^2 r^2}},
\end{equation}
which is positive when the Friedman equation, $k + \dot{a}^2 = \frac{8\pi}{3} \rho a^2$, is satisfied:
\begin{equation}
E_{\rm I}(\textbf{N}_{\rm Kodama}) = \frac{\frac{8 \pi}{3} \rho (a r)^3}{1 + \sqrt{1 - \frac{8 \pi}{3} \rho (a r)^2}} = \frac{2 m (\tilde{r})}{1 + \sqrt{1 - \frac{2 m (\tilde{r})}{\tilde{r}}}},
\end{equation}
where $m (\tilde{r}) = \frac{4 \pi}{3} \rho \tilde{r}^3$ and $\tilde{r} = a r$.

On the other hand, applying program II we get
\begin{equation}
E_{\rm II}(\textbf{N}_{\rm Kodama}) = \frac{l^2 (k + \dot{a}^2) a r^3}{\sqrt{l^2 + (\dot{a} r N^t + a N^r)^2} + \sqrt{1 - k r^2} N^t + \frac{a \dot{a} r}{\sqrt{1 - k r^2}} N^r}, \label{energy DB}
\end{equation}
where $l^2 = -g(\textbf{N}, \textbf{N})$. This energy is always positive for any timelike displacement vector $\textbf{N}$ when the Friedman equation, $k + \dot{a}^2 = \frac{8\pi}{3} \rho a^2$, is satisfied. Note that it vanishes for the Milne universe, i.e. $k = -1, a = t$, which is diffeomorphic to Minkowski spacetime.

For a comoving observer, the energy measured is
\begin{equation}
E_{\rm II}(\partial_t) = \frac{a r^3 (k + \dot{a}^2)}{\sqrt{1 + \dot{a}^2 r^2} + \sqrt{1 - k r^2}} = \frac{\frac{8 \pi}{3} \rho a^3 r^3}{\sqrt{1 + \dot{a}^2 r^2} + \sqrt{1 - k r^2}}.
\end{equation}
For a Kodama observer the result, after substituting $N^t, N^r$ from~(\ref{Kvec}), works out to be  the same as was found using program I:
\begin{equation}
E_{\rm II}(\textbf{N}_{\rm Kodama}) = E_{\rm I}(\textbf{N}_{\rm Kodama}).
\end{equation}

\section{An Alternative Approach}
We have seen that the proposed optimal programs give reasonable quasi-local energies: it is natural then to ask what are the determined references.
For simplicity let us consider the Schwarzschild geometry in the standard spherical coordinate and the unit future timelike displacement vector $\textbf{N}$. The embedding variables determined by program I are~\cite{Chen:2009zd,Wu:2011wk}
\begin{equation}
T_t = A N^t, \quad T_r = -A^{-1} N^r, \quad R_t = -N^r, \quad R_r = N^t.
\end{equation}
It is not very difficult to see that for $\textbf{N}$ to be the unit normal of the constant time hypersurface, i.e.,
\begin{eqnarray}
&& \textbf{N} = e_0 = \frac{1}{\sqrt{A}} \partial_t = \partial_T = \bar{e}_0
\nonumber\\
\Rightarrow \quad && N^t = \frac{1}{\sqrt{A}}, \qquad N^r = 0
\nonumber\\
\Rightarrow \quad && T_t = \sqrt{A}, \quad T_r = 0, \quad R_t = 0, \quad R_r = \frac{1}{\sqrt{A}},
\nonumber
\end{eqnarray}
the embedding isometrically matches the complete 4D orthonormal frame at the two-sphere boundary, this means the additional restrictions
\begin{eqnarray}
\bar{\vartheta}^0 &=& dT = T_t dt + T_r dr = \sqrt{A} dt = \vartheta^0,
\nonumber\\
\bar{\vartheta}^1 &=& dR = R_t dt + R_r dr = \frac{1}{\sqrt{A}} dr = \vartheta^1.
\nonumber
\end{eqnarray}

This is not extremely surprising since we are only considering spherically symmetric spacetimes here. So we propose an alternative approach here: the optimal choice of the reference should be the one which isometrically matches the 4D orthonormal frame at the two-sphere boundary, i.e., on the boundary we require
\begin{eqnarray}
\vartheta^0 = \bar{\vartheta}^0 &\qquad \Rightarrow \qquad& a_t dt + a_r dr = a_T dT + a_R dR,
\nonumber\\
\vartheta^1 = \bar{\vartheta}^1 &\Rightarrow& b_t dt + b_r dr = b_T dT + b_R dR, \label{isomatS}
\end{eqnarray}
which implies
\begin{eqnarray}
&& e_0 = \bar{e}_0, \qquad e_1 = \bar{e}_1, \qquad N^0 = N^{\bar{0}},
\nonumber\\
&& N^1 = N^{\bar{1}}, \qquad g(\textbf{N}, \textbf{N}) = \bar{g}(\textbf{N}, \textbf{N}),
\end{eqnarray}
and
\begin{eqnarray}
T_t = a_t b_R - a_R b_t, &\qquad& R_t = -a_t b_T + a_T b_t,
\nonumber\\
T_r = a_r b_R - a_R b_r, & & R_r = -a_r b_T + a_T b_r.
\end{eqnarray}
For these embeddings the second term of~(\ref{expB}) vanishes. By~(\ref{abtr}) we get the following results:
\begin{eqnarray}
X^{-1} &=& T_t R_r - T_r R_t
\nonumber\\
&=& (a_t b_R - a_R b_t) (-a_r b_t + a_T b_r) - (a_r b_R - a_R b_r) (-a_t b_T + a_T b_t)
\nonumber\\
&=& 1, \nonumber
\end{eqnarray}
and
\begin{eqnarray}
A N^t R_r + A^{-1} N^r R_t &=& -(A N^t a_r + A^{-1} N^r a_t) b_T + (A N^t b_r + A^{-1} N^r b_t) a_T
\nonumber\\
&=& -N^1 b_T + N^0 a_T = B N^T, \nonumber
\end{eqnarray}
so the energy obtained, for the RN--(A)dS metric, is
\begin{eqnarray}
E_{\rm iso} &=& r (B N^T - A N^t)
\nonumber\\
&=& r \left( \sqrt{l^2 B + (N^R)^2} - \sqrt{l^2 A + (N^r)^2} \right)
\nonumber\\
&=& \frac{l^2 (2 m - Q^2/r) + r ((N^R)^2 - (N^r)^2)}{\sqrt{l^2 B + (N^R)^2} + \sqrt{l^2 A + (N^r)^2}}. \label{ENTNt}
\end{eqnarray}
where $l^2 = -g(\textbf{N}, \textbf{N})$.

For any given displacement vector in the dynamic spacetime there is still one unknown in this energy expression, $N^T$ (or $N^R$). Physically to determine it means to pick a corresponding observer in the reference spacetime as the ground state. There are two obvious and physically meaningful choices we can think of, and, as we are going to show, they are equivalent to the energy-extremization programs I and II respectively. First, it is reasonable to require the reference displacement vector to be (proportional to) the timelike Killing vector, i.e., $\textbf{N} = N^T \partial_T$. Then~(\ref{ENTNt}) reduces to
\begin{equation}
E_{\rm isoI}(\textbf{N}) = r \left( \sqrt{-g(\textbf{N}, \textbf{N})B} - A N^t \right),
\end{equation}
which is equivalent to~(\ref{energy SA}). Second, it is also reasonable to pick a comparable reference observer as the ground state. By that we mean requiring, on the boundary,
\begin{equation}
\Lie_{\N} (\vartheta^2 \wedge \vartheta^3) = \Lie_{\N} (\bar{\vartheta}^{2} \wedge \bar{\vartheta}^{3}), \label{LNarea}
\end{equation}
where $\Lie_{\N}$ means the Lie derivative along the displacement vector $\textbf{N}$. This condition implies
\begin{equation}
N^R = N^r \qquad \Rightarrow \qquad B N^T = \sqrt{-g(\textbf{N}, \textbf{N}) B + (N^r)^2}
\end{equation}
so that~(\ref{ENTNt}) reduces to
\begin{equation}
E_{\rm isoII}(\textbf{N}) = r \left( \sqrt{-g(\textbf{N}, \textbf{N}) B + (N^r)^2} - A N^t \right),
\end{equation}
which is equivalent to~(\ref{energy SB}).

So one can see that this isometric matching approach gives results equivalent to the energy extremization programs. These two methods can be seen as complementary to each other. It helps us understand the geometric meaning of the energy-extremization program, at least for the spherically symmetric spacetimes, on the one hand. On the other hand, when we choose the reference that isometrically matches the 4D metric, the resultant energy would be the extremum.

Note that the quantities in the dynamic spacetime are known, and $N^{\bar{0}} = N^0$ and $N^{\bar{1}} = N^1$ after the exact matching~(\ref{isomatS}).
So from~(\ref{abtr}) the coefficients of the reference orthonormal coframes $(a_T, a_R, b_T, b_R)$ are functions of the embedding variables $(T_t, T_r, R_t, R_r)$, since
\begin{equation}
N^T = N^t T_t + N^r T_r, \qquad N^R = N^t R_t + N^r R_r.
\end{equation}
Then one can read~(\ref{aTbT}) as three constraints on $(T_t, T_r, R_t, R_r)$, so that now there is only one independent variable in~(\ref{ENTNt}). It is natural then to expect that one can reproduce $E_{\rm I}$ if one extremizes the energy~(\ref{ENTNt}) with respect to the single variable. Technically instead of the embedding variables $(T_t, T_r, R_t, R_r)$, it is more convenient to vary~(\ref{ENTNt}) with respect to the coefficients $(a_T, a_R, b_T, b_R)$. For simplicity we normalize the displacement vector,
\begin{equation}
g(\textbf{N}, \textbf{N}) = -1 = \bar{g}(\textbf{N}, \textbf{N}).
\end{equation}
From the equations for $a_T$ and $b_T$ in~(\ref{abtr}) we get
\begin{equation}
N^T = \frac{N^0 a_T - N^1 b_T}{B}. \label{NT}
\end{equation}
Now by extremizing~(\ref{ENTNt}) with respect to $a_T$ we get
\begin{equation}
0 = N^0 - N^1 \frac{\partial b_T}{\partial a_T},
\end{equation}
which together with~(\ref{aTbT}) gives
\begin{eqnarray}
&& \frac{\partial b_T}{\partial a_T} = \frac{a_T}{b_T}
\nonumber\\
\Rightarrow \quad && a_T = \sqrt{B} N^0, \quad b_T = \sqrt{B} N^1, \quad a_R = \frac{N^1}{\sqrt{B}}, \quad b_R = \frac{N^0}{\sqrt{B}},
\nonumber\\
\Rightarrow \quad && N^T = \frac{1}{\sqrt{B}}.
\end{eqnarray}
So the energy obtained is
\begin{equation}
E_{\rm isoex} = r (\sqrt{B} - A N^t) = E_{\rm I} = E_{\rm isoI}
\end{equation}
as expected.

Applying this isometric matching approach to the FLRW spacetime we get
\begin{equation}
E_{\rm iso} = a r \left( N^T - \sqrt{1 - k r^2} N^t - \frac{a \dot{a} r}{\sqrt{1 - k r^2} N^r} \right). \label{ENTNt D}
\end{equation}
If we require $\textbf{N} = N^T \partial_T$ then~(\ref{ENTNt D}) is equal to~(\ref{energy DA}). On the other hand, by requiring~(\ref{LNarea}), which implies $N^R = \dot{a} r N^t + a N^r$, the energy is the same as program II. If we extremize~(\ref{ENTNt D}) with respect to the only one independent variable, the result is the same as that determined by program I~(\ref{energy DA}).

Since we have explored so many different approaches to determine the reference and quasi-local energy, it may be suitable here to summarize these various approaches to quasi-local energies for spherically symmetric spacetimes by the Table~\ref{table}.

\begin{table}
\caption{The energies for the Reissner-Nordstr\"{o}m--(anti)-de Sitter and Friedman-Lema\^{\i}tre-Robertson-Walker spacetimes by different approaches.} \label{table}
\begin{tabular}[t]{cll}
\hline
 & \qquad energy for RN-(A)dS & \qquad energy for FLRW \\
\hline
iso   & $r (B N^T - A N^t)$
      & $a r (N^T - \frac{a N^t}{\tilde A} - \tilde A \dot{a} r N^r)$ \\
isoI  & $r (\sqrt{-g(\textbf{N}, \textbf{N}) B} - A N^t)$
      & $a r (\sqrt{l^2} - \frac{a N^t}{\tilde A} - \tilde A \dot{a} r N^r)$ \\
isoII & $r (\sqrt{-g(\textbf{N}, \textbf{N}) B + (N^r)^2} - A N^t)$
      & $a r (\sqrt{l^2 \!+\! (\dot{a} r N^t \!+\! a N^r)^2} - \frac{a N^t}{\tilde A} - \tilde A \dot{a} r N^r)$ \\
I     & $r (\sqrt{-g(\textbf{N}, \textbf{N}) B} - A N^t)$
      & $a r (\sqrt{l^2} - \frac{a N^t}{\tilde A} - \tilde A \dot{a} r N^r)$ \\
II     & $r (\sqrt{-g(\textbf{N}, \textbf{N}) B + (N^r)^2} - A N^t)$
      & $a r (\sqrt{l^2 \!+\! (\dot{a} r N^t \!+\! a N^r)^2} - \frac{a N^t}{\tilde A} - \tilde A \dot{a} r N^r)$ \\
\hline
     & $A = 1 - \frac{2m}{r} + \frac{Q^2}{r^2} - \frac{\Lambda}{3}r^2$ & $\tilde A = \frac{a}{\sqrt{1 - k r^2}}$ \\
     & $B = 1 - \frac{\Lambda}{3} r^2$ & $l^2 = -g(\textbf{N}, \textbf{N})$ \\
\hline
\end{tabular}
iso means given $(N^t, N^r)$ then matching the orthonormal frames. \\
isoI means iso with the restriction $N^R = 0$. \\
isoII means iso with the restriction \textbf{C1}. \\
I means given $(N^t, N^r)$, extremize the energy with no constraint. \\
II means given $(N^t, N^r)$, extremize the energy under the constraints \textbf{C1} and \textbf{C2}.
\end{table}

\section{Discussion}
As mentioned in the introduction, a principal issue in the covariant Hamiltonian formalism is the proper choice of reference spacetime for a given observer. When we embed the two-sphere boundary in the dynamic spacetime into the reference, the question becomes how to determine the embedding. In our earlier papers~\cite{Chen:2009zd,Wu:2011wk} we imposed the isometric embedding condition and minimized the energy with respect to the embedding variables with no further constraint, so the energy produced is the absolute minimum of all possible embeddings for any given displacement vector. For all different observers in the dynamic spacetime the program has chosen the same reference observer to be the ground state, so for some observers with extreme motion the energy measured can be extreme. In this paper we have proposed a modified program in which we minimize the energy under some additional suitable constraints. The main idea is to force the program to pick a comparable reference observer while minimizing the energy. Then no matter how extreme the motion is of an observer in the dynamic spacetime, he will be taking a comparable reference observer as the ground state. Then the results satisfy the usual criteria. The only exception is that for a small region inside the inner horizon of the Reissner-Nordstr\"{o}m spacetime the energy is negative. However, it is rather an odd region where the gravitational force becomes repulsive.

We have also shown that the 4D isometric matching approach produces compatible results. These two methods, energy-extremization and 4D isometric matching, can be seen as complementary to each other. It helps us understand the geometric meaning of the energy-extremization program, at least for the spherically symmetric spacetimes, on one hand. On the other hand, the most optimized references for spherically symmetric spacetimes, i.e., those that match the geometry on the boundary, produce the extremum quasi-local energies. We believe that this spherical case is the main test case; it shows that our programs have promise as universal approaches for determining the reference needed for the covariant Hamiltonian boundary term quasi-local energy for general spacetimes.

\section*{Acknowledgement}
We much appreciated several good suggestions from the referees which led to significant improvements.
This work was supported by the National Science Council of the R.O.C. under the grants NSC-99-2112-M-008-004, NSC-100-2119-M-008-018 (JMN) and NSC 99-2112-M-008-005-MY3 (CMC) and in part by the National Center of Theoretical Sciences (NCTS).



\end{document}